# Protoplanetary Earth Formation: Further Evidence and Geophysical Implications

J. Marvin Herndon

Transdyne Corporation
San Diego, California 92131 USA

August 30, 2004

## Abstract

Recently, I showed that the "standard model" of solar system formation is wrong, yielding the contradiction of terrestrial planets having insufficiently massive cores, and showed instead the consistency of Eucken's 1944 concept of planets raining out in the central regions of hot, gaseous protoplanets. Planets generally consist of concentric shells of matter, but there has been no adequate geophysical explanation to account for the Earth's non-contiguous crustal continental rock layer, except by assuming that the Earth in the distant past was smaller and subsequently expanded. Here, I show that formation of Earth, from within a Jupiter-like protoplanet, will account for the compression of the rocky Earth to about 64% of its current radius, yielding a closed, contiguous continental shell with concomitant Earth expansion commencing upon the subsequent removal of its protoplanetary gaseous shell. I now propose that Earth expansion progresses, not from spreading at mid-oceanic ridges as usually assumed, but primarily by the formation of expansion cracks (often near continental margins) and the in-filling of those cracks with basalt (produced from volume expansion in the mantle), which is extruded mainly at mid-oceanic ridges, solidifies and traverses the ocean floor by gravitational creep to regions of lower gravitational potential energy, ultimately plunging downward into distant expansion cracks, emulating subduction. Viewed from that perspective, most of the evidence presented in support of plate tectonics supports Earth expansion; mantle convection is not required, and the timescale for Earth expansion is no longer constrained to about 200 million years, the maximum age of the current ocean floor.

## Introduction

In 1944, on the basis of thermodynamic considerations, Arnold Eucken suggested core-formation in the Earth as a consequence of successive condensation from solar matter in the central region of a hot, gaseous protoplanet, with molten iron metal first raining out at the center (Eucken 1944). For a time, gaseous protoplanets were discussed (Kuiper 1951a; Kuiper 1951b; Urey 1952), but the idea of protoplanets was largely abandoned in favor of the so-called "standard model" of solar system formation, based upon the concept that grains condensed from diffuse nebula gases at a pressure of about $10^{-5}$ bar, and were then agglomerated into successively larger pebbles, rocks, planetesimals and, ultimately, planets. The problem, as I recently disclosed, is that the "standard model" of solar system formation would yield terrestrial planets having insufficiently massive cores, a profound contradiction to what is observed (Herndon 2004b). The "standard model" of solar system formation is *wrong* because the underlying "equilibrium condensation" model is *wrong*, the Earth in its composition is *not* like an ordinary chondrite meteorite as had been assumed, and condensates from nebula gases at $10^{-5}$ bar would be *too* oxidized to yield planetary cores of sufficient mass. Instead, within the framework of present knowledge, the concept of planets having formed by raining out from the central regions of hot, gaseous protoplanets, as revealed by Eucken (1944), appears to be consistent with the observational evidence (Herndon 2004b).

The Earth consists of more-or-less uniform, concentric shells of matter, except near the surface. There, units of the less-dense continental rock (sial) are separated by the denser ocean-floor basalt (sima). For more than a century, scientists have recognized that opposing margins of the continents fit together geographically and display geological and palaeobiological evidence of having in the past been joined (Hilgenberg 1933; Suess 1885; Wegener 1912). To date there has been no adequate geophysical explanation to account for the formation of the non-contiguous crustal continental rock layer, except by assuming that the Earth in the distant past had a smaller volume and, consequently, had a smaller surface area (Carey 1976; Hilgenberg 1933; Scalera & Jacob 2003). In the following I show that formation of our home planet from within a Jupiter-like protoplanet (Herndon 2004b) will account for the compression of the Earth to a dimension such that the continental sial shell is uniform and continuous. The geophysical implications of the rocky Earth's subsequent expansion, following removal of its protoplanetary gaseous shell, are outlined briefly, and will be discussed in greater detail in a subsequent communication.

## Protoplanetary Compression of the Earth and Subsequent Expansion

Previously, I have shown from fundamental ratios of mass that the Earth is like a highly reduced enstatite chondrite meteorite and *not* like an ordinary chondrite. I have demonstrated: (*i*) that the Earth has a state of oxidation like an enstatite chondrite (Herndon 1996; Herndon 2004a), (*ii*) that the components of the endo-Earth, the inner 82% comprising the lower mantle and core, are quite similar to the components of a particular enstatite chondrite meteorite (Herndon 1980; Herndon 1993); the upper mantle appears to contain additional, oxidized, and undifferentiated components, suggestive of layers of veneer (Herndon 1998), (*iii*) that the relative mass of the core is an inverse function of planetary oxygen content (Herndon & Suess 1977), and (*iv*) that the composition of the core itself is a function of oxidation state (Herndon 1993).

On the basis of thermodynamic considerations, Hans Suess and I showed that some of the minerals of enstatite meteorites could form at high temperatures in a gas of solar composition at pressures above about 1 bar, provided thermodynamic equilibria are frozen in at near-formation temperatures (Herndon & Suess 1976). At such pressures, molten iron, together with the elements that dissolve in it, is the most refractory condensate. Although there is much to verify and learn about the process of condensation from near the triple point of solar matter, the glimpses we have seen are remarkably similar to vision of Eucken (1944), *i.e.*, molten iron raining out in the center of a hot, gaseous protoplanet.

The mass of protoplanetary-Earth, about 275 to 305 times the mass of the present-day Earth, a value which is quite similar to Jupiter's mass, $318 m_E$, may be calculated from solar abundance data (Anders & Grevesse 1989) by adding to the condensable, planetary elements their proportionate amount of solar elements that are typically gases (*e.g.*, H, He) or form volatile compounds (*e.g.*, O, C, N). To emphasize the similarity with Jupiter, the principal primordial constituents of protoplanetary Earth are expressed in Table 1 in terms of major Jovian components. Under such a great overburden of gases, the relatively non-volatile alloy-plus-rock constituents, which now comprise most of the Earth, would be compressed by gravity to a fraction of the present day Earth radius.

Pressures at the gas-rock boundary within the interior of Jupiter are estimated to be in the range from 43 Mbar to 60 Mbar (Podolak & Cameron 1974; Stevenson & Salpeter 1976). At the pressures of interest, density becomes a function almost entirely of atomic number and atomic mass (March 1957). Using a theoretical Thomas-Fermi-Dirac approach, in which errors are thought to decrease as pressure increases (Stevenson & Salpeter 1976), I calculated density at Jupiter-model, gas-rock-boundary pressures for matter having the approximate composition of our Earth as a whole. The calculations are based upon eight chemical elements that account about 98% of the Earth's mass, assume volume additivity, and ignore phase separations and transitions. The results of the calculations, presented in Table 2, show that a Jovian-like gas envelope is sufficient to compress the protoplanetary alloy-plus-rock core that became the Earth to an average density of 21 g/cm$^3$ which, as shown below, is consistent with the density required for a contiguous, closed, crustal continental sial shell prior to Earth expansion.

Approximately 29% of the surface area of the Earth is composed of the portions of continents that presently lie above mean sea level; an additional 12% of the surface area of the Earth is composed of the continental margins, which are submerged to depths of no more than 2 km (Mc Lennan 1991). Notably, the margins of the continents, so defined, appear to fit considerably better together, as pieces of a jigsaw puzzle, than when defined by the coastlines (Carey 1976). The continents, including the continental margins, comprise approximately 41% of the present surface area of the Earth. Compression of the Earth to a radius of 4077 km, *i.e.*, 64% of its currant radius, would result in the continents becoming a uniform, continuous crustal sial shell with a mean density for the Earth of 21 g/cm$^3$. The density value of 21 g/cm$^3$ is identical to that estimated to result from protoplanetary compression by the great mass of Jovian-like gases (Table 2) and, I submit, stands as further evidence of the formation of Earth from within a giant, gaseous protoplanet (Herndon 2004b).

Decompression and volume expansion of the Earth may be seen as a direct consequence of the subsequent removal of hydrogen and other volatile constituents, presumably by the high temperatures and/or by the violent activity during some early super-luminous solar event, such as the T-Tauri phase solar wind associated with the thermonuclear ignition of the Sun. After being stripped of such a great overburden, the Earth would rebound, tending toward a new hydrostatic equilibrium by expanding. Gravitational energy of compression, stored during the Jupiter-like protoplanetary stage, may be seen as the primary energy source for geodynamic activity related to the expansion of the Earth during its approach toward a new hydrostatic equilibrium. To a much lesser extent, nuclear fission energy (Herndon 1994; Herndon 2003; Hollenbach & Herndon 2001) and radioactive decay energy may augment the stored energy of protoplanetary compression, heating the interior of the Earth, and compensating to some extent for the cooling that results from expansion.

## Geophysical Implications

Generally, planets consist of concentric shells of matter. Our Earth with its non-contiguous continent distribution would seem to be an exception, inexplicable unless at some past time its surface area was markedly less than today's. Because the sial continental crustal rock is less dense, portions of that matter cannot have sunk into the ocean floors or disappeared into the interior by subduction. Many geoscientists have devoted considerable efforts to find ways to explain how the Earth might have been smaller in the past and a great many ideas have been put forth related to various possible geodynamic mechanisms that might be involved in Earth expansion. For references, see (Scalera & Jacob 2003), especially the superb bibliographic database contained therein.

Geo-tectonic ideas stemming from Earth expansion pre-date those of plate tectonics and until now there has been little ground for common agreement. Perhaps the chief cause for that disparity is that each (as currently formulated) is not quite correct. For example, plate tectonics appears to describe well certain ocean-floor observations, such as magnetic striations, but there are some concerns as to whether concomitant mantle convection and the recycling of basalt can and/or actually does take place (Yoder 1990). Likewise, Earth expansion theory, as espoused by Carey (1988) and others, is predicated upon the idea that Earth expansion occurs mainly along mid-oceanic ridges during the last 200 million years, as the oldest ocean floor is no older than that (Khramov 1982).

Viewed from a different perspective, however, it may be possible to reconcile the two seemingly divergent theories, plate tectonics and Earth expansion, into one unified theory. I now propose that Earth expansion progresses, not from spreading at mid-oceanic ridges, but primarily by the formation of expansion cracks (often near continental margins) and the in-filling of those cracks with basalt (produced from volume expansion in the mantle), which is extruded mainly at mid-oceanic ridges, solidifies and traverses the ocean floor by gravitational creep to regions of lower gravitational potential energy, ultimately plunging downward into distant expansion cracks, emulating subduction. Viewed from that perspective, most of the evidence presented in support of plate tectonics supports Earth expansion; mantle convection is not required, and the timescale for Earth expansion is no longer constrained to 200 million years, the maximum age of the current ocean floor.

The major obstacle to acceptance of Earth expansion, as envisioned by Hilgenberg (1933), Carey (1988) and others, has been that scientists have lacked knowledge of a mechanism that could provide the necessary quantity of energy for such a great expansion of the Earth (Beck 1961; Cook & Eardley 1961) without departing from the physical laws of nature as presently understood (Jordan 1971). Scheidegger (1982) stated concisely the prevailing view, "Thus, if expansion on the postulated scale occurred at all, a completely unknown energy source must be found." Here, I have disclosed that energy source -- protoplanetary energy of compression. There are yet fundamental questions that need to be addressed (particularly those related to timescale). And they should be addressed, not by making models based upon arbitrary assumptions, but rather, by making discoveries and by discovering fundamental quantitative relationships in nature.

**Table 1.** Comparison of the composition of protoplanetary Earth with that of Jupiter, calculated from solar elemental abundance data (Anders & Grevesse 1989), expressed in units of Earth masses ($m_E=1$), shown as major chemical compounds common to Jupiter, with masses of Jupiter and Saturn shown for comparison.

| Chemical Substance | Mass Relative to Earth Mass |
|---|---|
| hydrogen | 219 |
| inert gases | 80 |
| water | 3.0 |
| methane | 1.3 |
| ammonia | 0.4 |
| alloy and rock (Earth) | 1.0 |
| total | 305 |
| Jupiter | 318 |
| Saturn | 95 |

**Table 2.** Published model pressure and density estimates (Podolak & Cameron 1974; Stevenson & Salpeter 1976) at the gas-rock boundary of Jupiter, shown for comparison with theoretical calculation of compressed Earth density at the same pressures.

| Jupiter Model Pressure (Mbar) | Jupiter Model Density (g/cm$^3$) | Compressed Earth Density (g/cm$^3$) |
|---|---|---|
| 43 | 18 | 20 |
| 46 | 18 | 21 |
| 60 | 20 | 23 |